\documentclass[11pt]{article}

\usepackage[left=1.3in,right=1.3in,top=1in,bottom=1in]{geometry}
\usepackage{graphicx}   
\usepackage{amsmath,amssymb}
\usepackage{cite}       
\usepackage{hyperref}   
\usepackage{caption}

\title{Self-optimizing multichannel optical computing}
\author{
Fatma Nur K{\i}l{\i}n\c{c}$^{1}$ and U\u{g}ur Te\u{g}in$^{1,2,*}$ \\
\small $^{1}$Department of Computational Sciences and Engineering, Ko\c{c} University, \.{I}stanbul, T\"urkiye \\
\small $^{2}$Department of Electrical and Electronics Engineering, Ko\c{c} University, \.{I}stanbul, T\"urkiye \\
\small $^{*}$Corresponding author: utegin@ku.edu.tr
}
\date{} 

\begin{document}
\maketitle

\begin{abstract}
Optical computing offers ultrafast, energy-efficient alternatives to conventional digital processors, yet most implementations remain confined to single-channel processing, severely underutilizing light's information capacity. Here we demonstrate a self-optimizing multichannel optical computing architecture based on multi-plane light conversion that natively processes RGB images and structured numerical data throughout the optical domain. We introduce two complementary optimization strategies that enable autonomous performance adaptation without differentiable forward models. First, Bayesian optimization tunes channel mixing coefficients to minimize crosstalk and enhance feature separability at the input level. Second, a hardware-in-the-loop protocol based on self-organized criticality leverages avalanche dynamics to autonomously navigate the high-dimensional phase landscape, enabling the system to self-optimize through stochastic multi-scale perturbations. Across medical imaging, natural image classification, and regression tasks, multichannel processing with random phase masks improves accuracy by 26--58 percentage points over raw pixel baselines, with RGB systematically outperforming grayscale by 5--6 percentage points. Self-optimization strategies provide additional gains of 6--7 percentage points through autonomous adaptation at complementary system levels. Our work establishes self-optimizing multichannel optical computing as a practical platform for real-world machine learning applications.
\end{abstract}

\section{Introduction}

The relentless growth of data volumes and computational demands in artificial intelligence and scientific computing has exposed fundamental bottlenecks in conventional digital hardware. Modern processors, constrained by the von Neumann architecture, suffer from memory-access latency, limited parallelism, and escalating power consumption, particularly as model sizes expand~\cite{ref1, ref2}. The energy footprint of training and deploying large-scale machine learning models now rivals that of entire data centers~\cite{ref3}, prompting an urgent search for alternative computing paradigms. Among these, optical computing has resurged as a powerful candidate, offering ultrafast signal propagation, intrinsic parallelism, and energy efficiency by exploiting the physics of light itself rather than charge transport~\cite{ref4, ref5}. Optical systems can natively perform key mathematical primitives---such as Fourier transforms, convolutions, and matrix multiplications---in a single step using passive or minimally powered elements~\cite{ref6, ref7}.

Over the past decade, photonic neural networks and analog optical computing platforms have demonstrated remarkable progress across diverse architectures, including free-space diffractive systems~\cite{ref8, ref9, ref10, ref11, ref12, ref13, ref14}, nonlinear multimode fibers~\cite{ref15, ref16}, and nanophotonic circuits~\cite{ref17, ref18, ref19}. These implementations exploit light--matter interactions to transform and process high-dimensional data at the speed of light, with recent advances in spatiotemporal chaos and nonlinearity~\cite{ref20}, supercontinuum generation~\cite{ref21}, and programmable multimode fiber propagation~\cite{ref22} pushing performance to new frontiers. Yet, a critical limitation persists: most existing optical computing approaches remain confined to single-channel processing~\cite{ref31}. Even when handling inherently multichannel or color datasets, information is often collapsed into a single grayscale representation before optical computation. This severely underutilizes the full information capacity of light---specifically its spectral dimension---and restricts scalability for modern machine learning tasks. Recent work has begun to address this gap through multichannel optical neural network architectures that exploit interference and diffraction for inter- and intra-channel processing~\cite{ref31, ref32}, yet these approaches have not been combined with gradient-free, hardware-in-the-loop optimization strategies that are essential for practical deployment.

Recent advances in multi-plane light conversion (MPLC) have demonstrated the capability to perform complex transformations on optical fields~\cite{ref23, ref24}. MPLCs offer a promising route to telecommunication~\cite{ref25, ref26}, quantum optics~\cite{ref27, ref28}, and hardware-accelerated machine learning~\cite{ref29, ref30}. In these systems, cascaded nonlinear or random phase modulations can increase feature diversity and emulate aspects of deep learning models entirely with light. However, current implementations largely target binary or single-channel classification tasks and struggle to accommodate multichannel data (such as RGB images) or structured numerical inputs typical of regression problems. Additionally, efficiently encoding diverse inputs and mitigating crosstalk remain open challenges, particularly when processing multiplexed information in real-world optical devices. Moreover, the optimization of high-dimensional phase-mask configurations typically relies on gradient-based methods that require accurate differentiable models of wave propagation~\cite{ref11}, which are sensitive to device nonidealities, calibration drift, and model--hardware mismatch~\cite{ref37}. Recent work has addressed this through transmission-matrix-based self-configuration~\cite{ref37}, though such approaches require complex interferometric measurement protocols and bidirectional optical access. Alternative gradient-free metaheuristics, such as evolutionary algorithms, remain computationally demanding and require careful hyperparameter tuning~\cite{ref14}. Self-organized criticality (SOC)---a phenomenon observed in complex systems that naturally generates multi-scale, spatially correlated perturbations---has recently emerged as a promising optimization paradigm for non-convex landscapes~\cite{ref33,PhysRevLett.59.381}, requiring neither differentiable models nor complex measurement protocols. SOC has demonstrated efficient search capabilities in combinatorial optimization and has been proposed as a mechanism underlying adaptive information processing in neural systems~\cite{ref34, ref35, ref36}, yet its application to optical computing hardware optimization remains unexplored.

Here, we report a self-optimizing programmable multi-plane light-conversion (MPLC) architecture designed specifically for multichannel machine learning. A phase-only spatial light modulator (SLM) encodes diverse input modalities in a spatially multiplexed format, while a cascade of phase-only masks implements the optical transformation. After intensity detection, the resulting patterns provide nonlinear features that are mapped to labels by a simple readout layer. To maximize performance, we introduce two distinct self-optimization strategies that operate autonomously without manual parameter tuning. First, for input optimization, we employ Bayesian optimization to fine-tune channel mixing coefficients, thereby minimizing crosstalk and enhancing feature separability at the source. Second, we implement a hardware-in-the-loop training protocol based on self-organized criticality (SOC). This approach leverages avalanche dynamics from an Abelian sandpile model to autonomously navigate the non-convex phase landscape, enabling the system to self-optimize without the need for differentiable forward models or precise physical alignment. We demonstrate the versatility of our architecture across a broad spectrum of tasks, including regression on structured numerical data, medical image classification, and natural image object recognition, establishing self-optimizing multichannel optical computing as a practical platform for real-world machine learning applications.

\section{Results}

\subsection{Multichannel MPLC architecture for optical machine learning}

\begin{figure}[ht!]
\centering
\includegraphics[width=\textwidth]{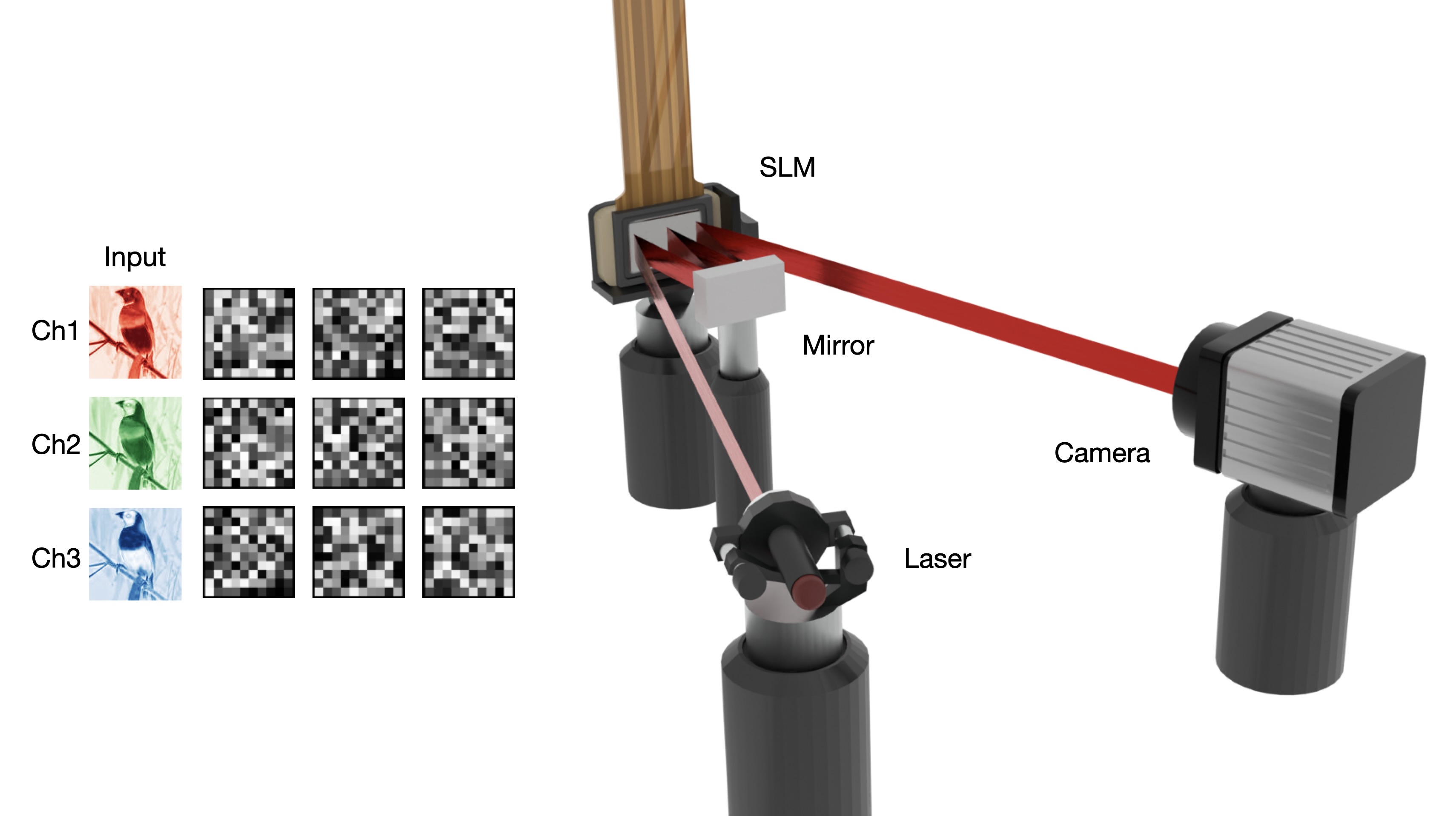}
\caption{\textbf{Multichannel optical computing architecture and concept.} (\textbf{a}) Experimental setup showing the spatial light modulator (SLM), planar mirror, and camera detection system. The laser beam undergoes multiple reflections to create cascaded phase transformations. (\textbf{b}) Conceptual diagram illustrating multichannel input encoding (RGB channels spatially multiplexed), phase mask cascade, and intensity detection followed by digital readout. (\textbf{c}) Example phase patterns displayed on the SLM active regions, demonstrating the programmable nature of the optical transformation.}
\label{fig:setup}
\end{figure}

To exploit the full information capacity of light, we designed an MPLC system that natively processes multichannel inputs without collapsing them to grayscale (Fig.~\ref{fig:setup}). The architecture comprises three functional components: input encoding, optical transformation, and feature extraction.

In the input stage, diverse data modalities---RGB image channels, structured numerical features, or other vectorized inputs---are spatially encoded onto distinct regions of a phase-only spatial light modulator. The encoded optical field then propagates through a cascade of phase-only masks implemented via multiple reflections between the SLM and a planar mirror, creating spatially separated active regions. Each mask--propagation pair implements a unitary transformation, and their composition realizes a multi-layer optical network. Crucially, the cascade preserves multichannel structure throughout propagation---the channels interact through interference and diffraction, generating feature-rich intensity patterns that encode both intra-channel structure and inter-channel correlations. At the detection plane, a camera captures the intensity distribution, yielding feature vectors that are mapped to task outputs via a simple digital readout layer.

This architecture offers two key advantages over prior optical computing approaches. First, by maintaining multichannel encoding throughout the optical domain, it fully exploits the dimensionality of the input data---particularly critical for RGB images where 67\% of information would be discarded by grayscale conversion. Second, the phase-only implementation ensures lossless, energy-efficient propagation while retaining full programmability for task-specific optimization. Critically, both the input encoding (via channel mixing) and the optical transformation (via phase masks) can be autonomously optimized, enabling self-adapting performance without manual parameter tuning or differentiable models. We evaluate this architecture across classification and regression benchmarks, demonstrating that multichannel optical processing provides substantial performance gains, and that these gains can be further enhanced through complementary self-optimization strategies operating at the input (channel mixing) and hardware (phase mask) levels.

\subsection{Multichannel optical computing outperforms single-channel processing}

We first establish the fundamental advantage of multichannel over single-channel optical computing by comparing grayscale and RGB processing on medical imaging and natural image classification tasks. All experiments in this section use randomly initialized phase masks and the same digital readout (Ridge classifier), isolating the contribution of multichannel encoding.

\begin{figure}[ht!]
\centering
\includegraphics[width=\textwidth]{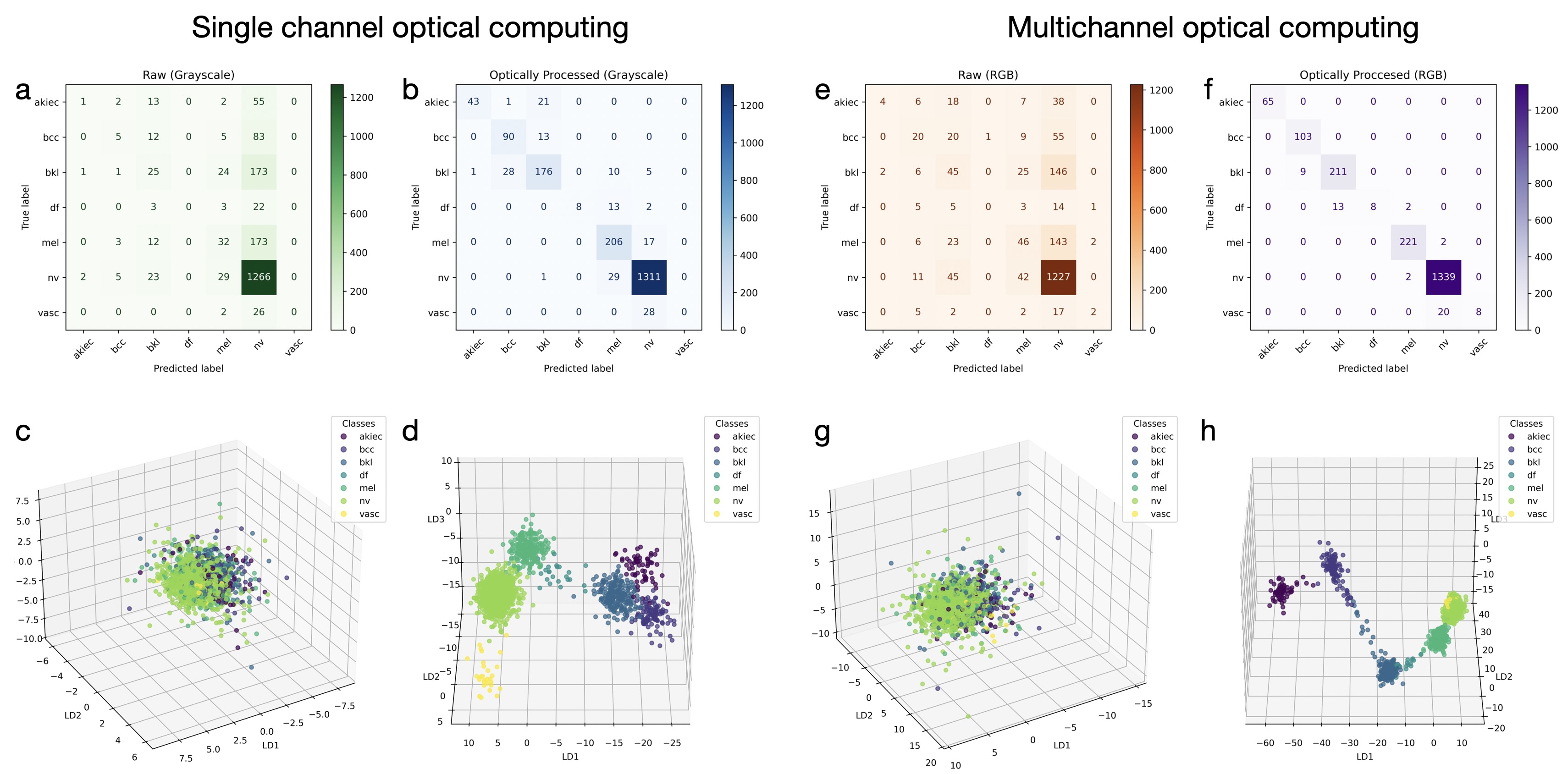}
\caption{\textbf{Single-channel and multichannel optical computing for HAM10000 dermatoscopic lesion classification.} (\textbf{a}) Confusion matrix for raw grayscale inputs. (\textbf{b}) Confusion matrix for optically processed grayscale inputs. (\textbf{c}) LDA projection showing raw grayscale feature space. (\textbf{d}) LDA projection showing optically processed grayscale features with improved cluster separation. (\textbf{e}) Confusion matrix for raw RGB inputs. (\textbf{f}) Confusion matrix for optically processed RGB inputs, showing near-perfect classification. (\textbf{g}) LDA projection for raw RGB features. (\textbf{h}) LDA projection for optically processed RGB features, demonstrating tight, well-separated clusters across all seven diagnostic categories.}
\label{fig:ham10000}
\end{figure}

\textbf{HAM10000 dermatoscopic lesion classification.} On the HAM10000 dataset (10,015 dermoscopic images across 7 diagnostic categories), multichannel optical processing dramatically improves classification accuracy (Fig.~\ref{fig:ham10000}, Table~\ref{tab:performance_all}). For grayscale inputs, raw pixel features achieve 66\% accuracy with the Ridge classifier, while optical processing through the phase-mask cascade increases accuracy to 92\% (+26 percentage points). For RGB inputs, raw features achieve 67\% accuracy, and multichannel optical processing reaches 98\% (+31 percentage points). Confusion matrices (Fig.~\ref{fig:ham10000}a,b,e,f) show substantially improved diagonal dominance after optical processing for both modalities, indicating reduced cross-class confusion. Linear discriminant analysis (LDA) projections (Fig.~\ref{fig:ham10000}c,d,g,h) reveal that optical features form tighter, more separable clusters compared to the overlapping distributions of raw pixel features. Notably, RGB processing outperforms grayscale by 6 percentage points (98\% vs.\ 92\%), demonstrating that preserving color information through the optical transformation enhances discriminability for this medical imaging task where subtle color variations carry diagnostic significance.

\begin{figure}[ht!]
\centering
\includegraphics[width=\textwidth]{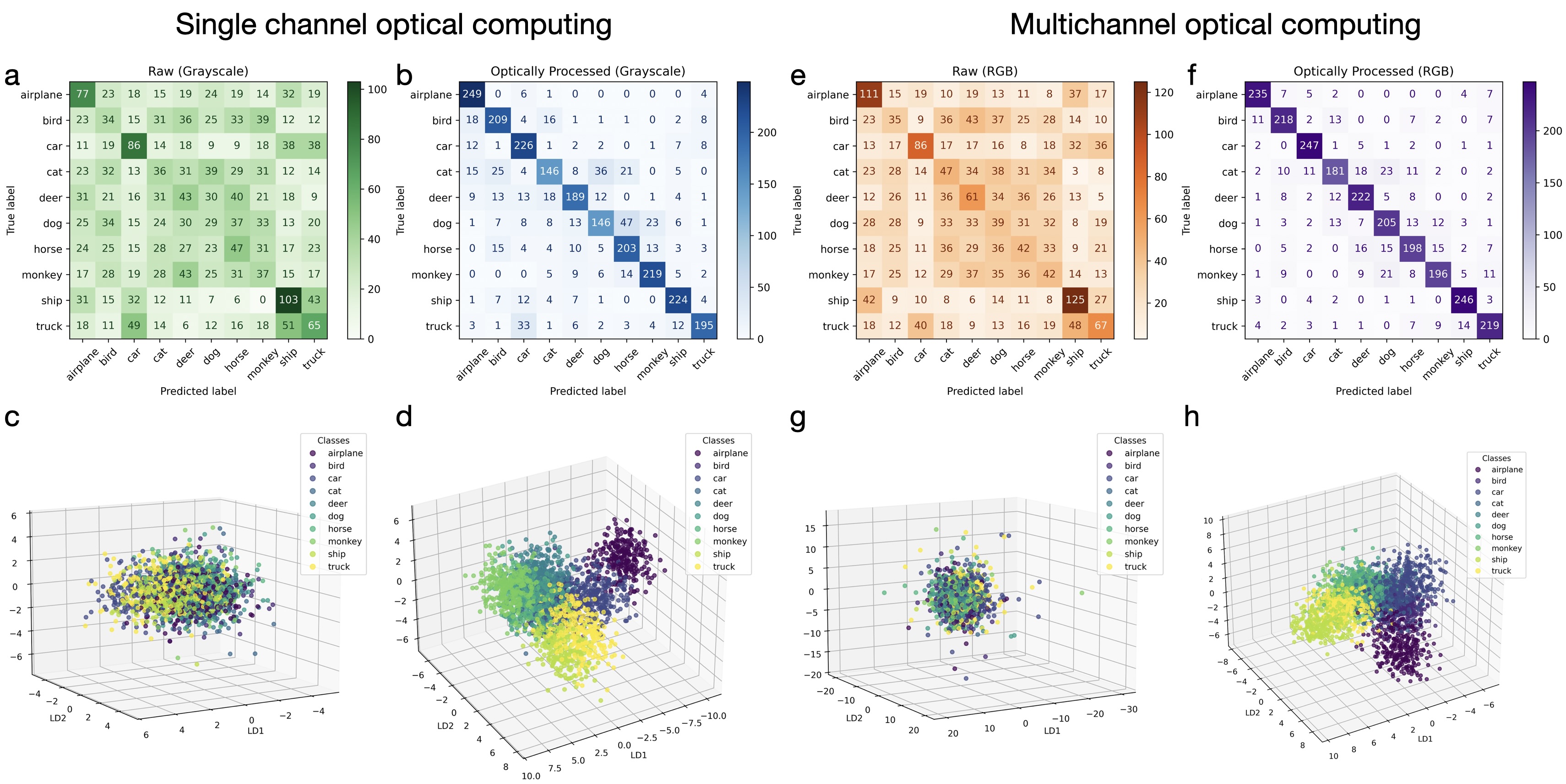}
\caption{\textbf{Single-channel and multichannel optical computing for STL-10 natural image classification.} (\textbf{a}) Confusion matrix for raw grayscale inputs showing poor discrimination across object categories. (\textbf{b}) Confusion matrix for optically processed grayscale inputs with substantial improvement. (\textbf{c}) LDA projection for raw grayscale features showing heavily overlapping clusters. (\textbf{d}) LDA projection for optically processed grayscale features with improved separation. (\textbf{e}) Confusion matrix for raw RGB inputs. (\textbf{f}) Confusion matrix for optically processed RGB inputs demonstrating strong category discrimination. (\textbf{g}) LDA projection for raw RGB features. (\textbf{h}) LDA projection for optically processed RGB features showing well-defined, separated clusters for all ten object categories.}
\label{fig:stl10}
\end{figure}

\textbf{STL-10 natural image recognition.} To assess generalization beyond medical imaging, we evaluated the system on STL-10 (13,000 natural images across 10 object categories). Multichannel optical processing provides even larger improvements on this more challenging dataset (Fig.~\ref{fig:stl10}, Table~\ref{tab:performance_all}). Grayscale processing increases accuracy from 21\% (raw pixels) to 77\% (+56 percentage points), while RGB processing improves from 25\% to 83\% (+58 percentage points). Confusion matrices (Fig.~\ref{fig:stl10}a,b,e,f) and LDA projections (Fig.~\ref{fig:stl10}c,d,g,h) again confirm the progression from poorly separated raw features to well-structured optical feature spaces. The RGB advantage over grayscale (83\% vs.\ 77\%, +6 percentage points) remains consistent, indicating that multichannel encoding provides robust benefits across diverse image domains.

\textbf{Abalone age regression.} To demonstrate that multichannel optical computing extends beyond image modalities, we applied the same optical pipeline to structured numerical data (Abalone dataset, 4,177 samples). After preprocessing (normalization and one-hot encoding of categorical variables), nine features were spatially encoded as a $3\times3$ grid on the SLM. The resulting optical features, processed by a Ridge regressor, achieve a normalized root mean square error of 0.08 (Table~\ref{tab:performance_all}; see Methods for normalization). This result confirms that the proposed architecture functions as a general-purpose optical feature extractor for multichannel data, with appropriate spatial encoding enabling effective processing of non-image structured inputs.

\begin{figure}[ht]
\centering
\includegraphics[width=0.95\textwidth]{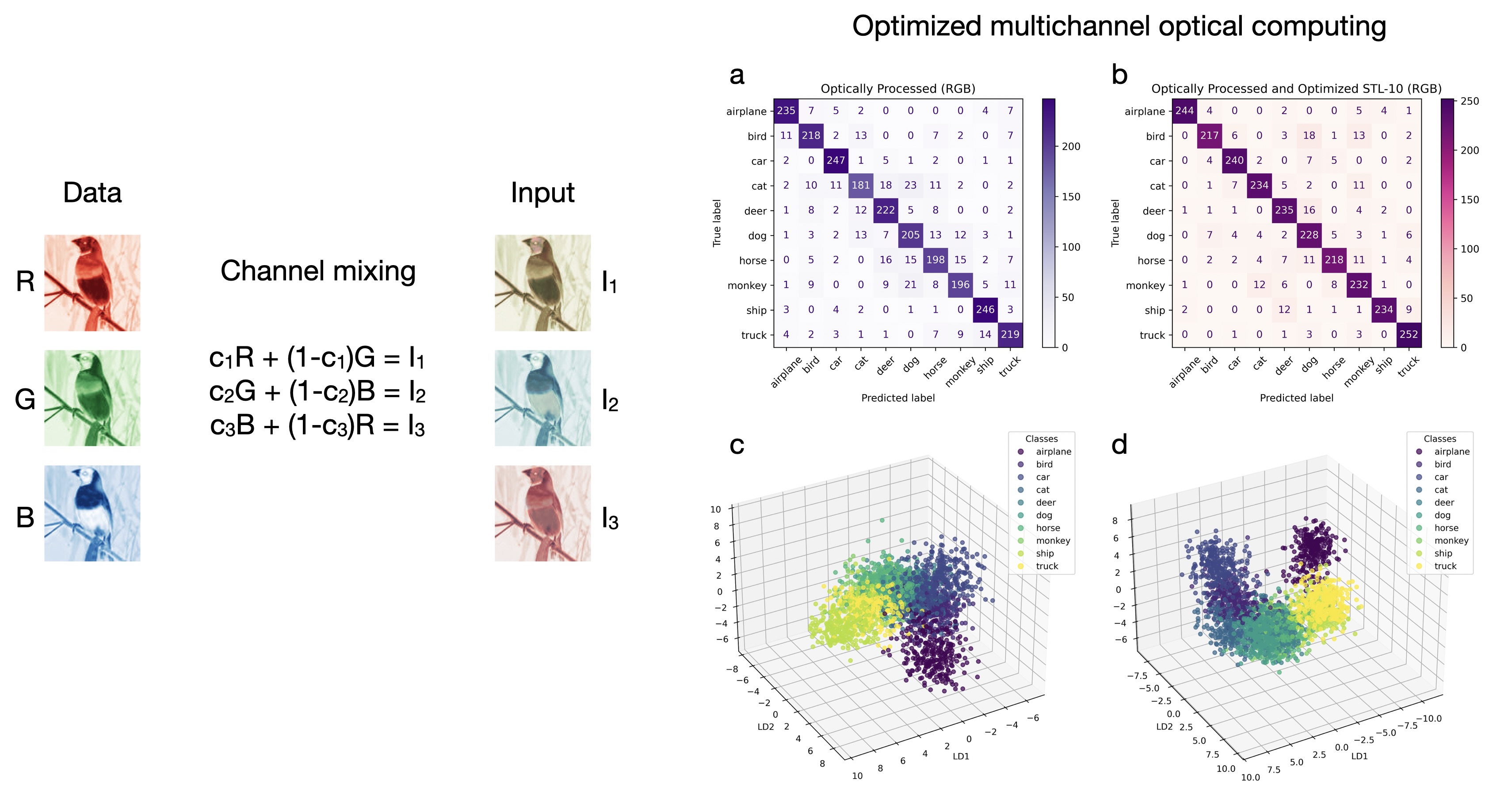}
\caption{\textbf{Input-space optimization via Bayesian channel mixing.} Left: Schematic showing multichannel RGB data (R, G, B) undergoing learned linear mixing with optimized coefficients before optical encoding. (\textbf{a}) Confusion matrix for optically processed RGB inputs without channel mixing. (\textbf{b}) Confusion matrix after Bayesian optimization of mixing coefficients, showing improved diagonal dominance and reduced confusion between similar categories. (\textbf{c}) LDA projection for unmixed RGB optical features. (\textbf{d}) LDA projection after channel mixing optimization, demonstrating tighter, more separated clusters across all ten STL-10 categories.}
\label{fig:channel_mixing}
\end{figure}

\begin{figure}[ht!]
\centering
\includegraphics[width=0.65\textwidth]{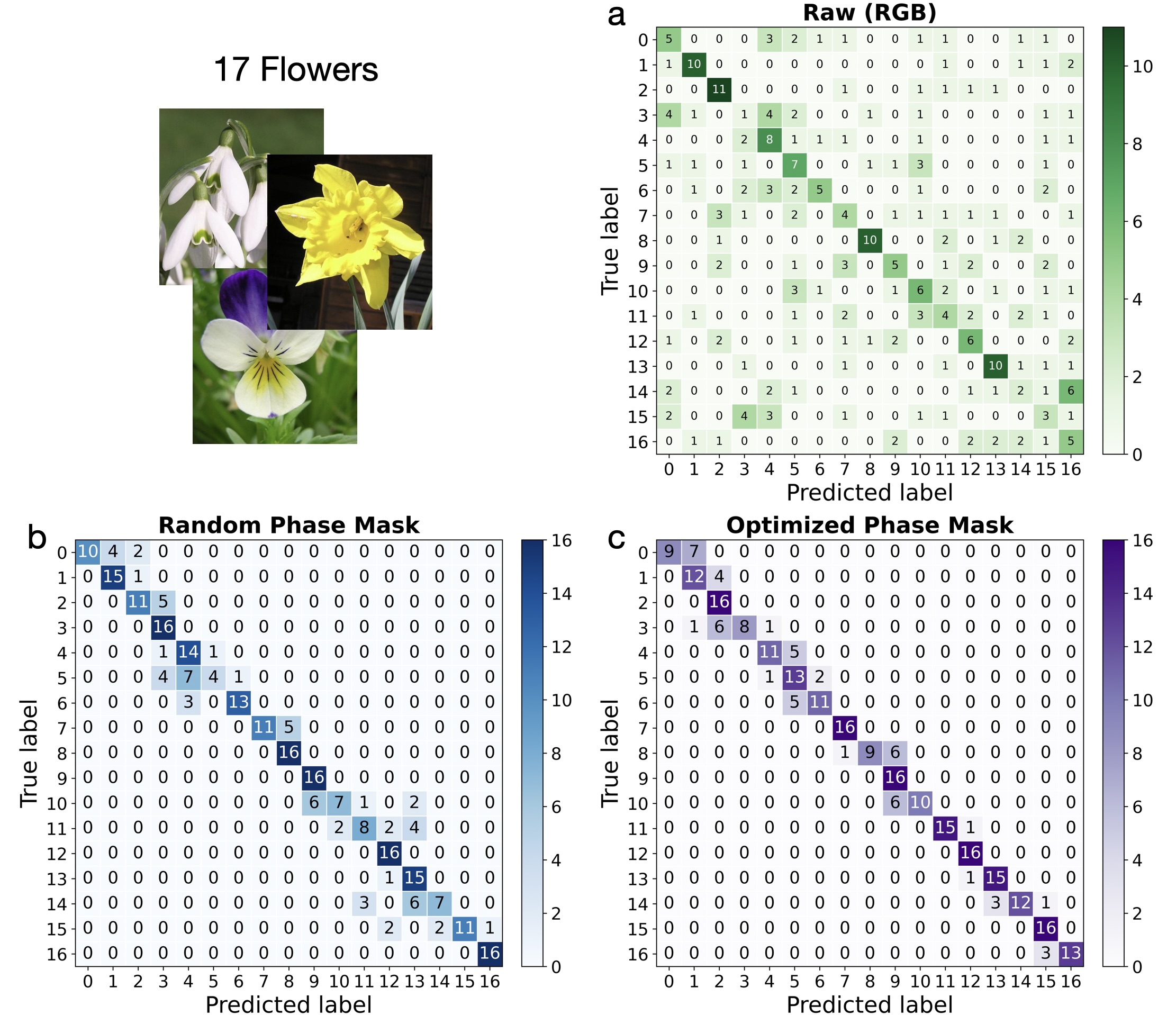}
\caption{\textbf{Hardware-space optimization via self-organized criticality for Oxford Flowers-17 classification.} Representative flower images from the dataset are shown at top. (\textbf{a}) Confusion matrix for raw RGB pixel inputs processed by Ridge classifier (37\% accuracy), showing poor discrimination across the 17 flower categories. (\textbf{b}) Confusion matrix after optical processing with random phase masks (74\% accuracy), demonstrating substantial improvement from optical feature extraction alone. (\textbf{c}) Confusion matrix after SOC-based phase mask optimization (80\% accuracy), showing further refinement with particularly strong diagonal dominance and reduced inter-class confusion.}
\label{fig:soc}
\end{figure}

\subsection{Input-space optimization via Bayesian channel mixing}

While multichannel encoding preserves information content, spatial multiplexing can introduce inter-channel interference and suboptimal utilization of available optical degrees of freedom. To address this, we introduce a channel mixing module that applies learned linear combinations to input channels before optical encoding. We parameterize mixing with a small set of coefficients and optimize them via Bayesian search while keeping phase masks fixed at their random initialization.

On STL-10 RGB classification, Bayesian-optimized channel mixing improves accuracy from 83\% (unmixed RGB with random masks) to 90\% (+7 percentage points; Fig.~\ref{fig:channel_mixing}, Table~\ref{tab:performance_all}). Confusion matrices (Fig.~\ref{fig:channel_mixing}a,b) show enhanced discriminability across all categories, with particularly strong improvements for visually similar classes. LDA projections (Fig.~\ref{fig:channel_mixing}c,d) confirm that optimized mixing produces more compact and separated feature clusters. Critically, this performance gain is achieved purely through software-level tuning---the optical hardware and phase masks remain unchanged. This input-space self-optimization provides a practical, low-cost path to improved multichannel performance without physical reconfiguration of the optical system.

\subsection{Hardware-space optimization via self-organized criticality}

We next optimize the phase masks directly under hardware-in-the-loop (HIL) constraints using self-organized criticality. Unlike gradient-based methods that require differentiable forward models or in-situ gradient measurements, SOC leverages avalanche dynamics from an Abelian sandpile model to generate spatially correlated, multi-scale perturbations. This enables autonomous exploration of the high-dimensional phase landscape without hyperparameter tuning or model assumptions. We disable channel mixing in these experiments to isolate the contribution of mask optimization, and select Oxford Flowers-17 as a practical testbed that balances inter-class variability with manageable sample size for iterative HIL evaluation.

Starting from random phase mask initialization (74\% accuracy), SOC-based optimization improves accuracy to 80\% within approximately 75 iterations (+6 percentage points; Fig.~\ref{fig:soc}, Table~\ref{tab:performance_all}). Confusion matrices (Fig.~\ref{fig:soc}a,b,c) show progressive improvement in classification performance, with particularly strong gains for flower categories that were initially confused. The rapid convergence---achieving meaningful improvements within hours of HIL training---demonstrates that SOC efficiently navigates the non-convex optimization landscape despite lacking gradient information. Raw pixel inputs without optical processing achieve only 37\% accuracy on this dataset, confirming that both the random optical transformation (74\%) and SOC optimization (80\%) provide substantial value. Notably, SOC optimization complements the channel mixing strategy: Bayesian optimization tunes input encoding with fixed hardware (Section~2.3), while SOC adapts the optical transformation itself when mask updates are feasible.

\section{Discussion}

We have demonstrated a self-optimizing multichannel optical computing architecture that addresses two fundamental limitations of prior work: the inability to natively process multichannel data throughout the optical domain, and the reliance on gradient-based optimization methods that require accurate differentiable models of the physical system. By preserving RGB or other multichannel structure during propagation, our system achieves substantial performance improvements over single-channel alternatives across medical imaging, natural image classification, and structured data regression. Two complementary self-optimization strategies---Bayesian search over input mixing coefficients and self-organized criticality for phase mask optimization---each provide additional gains and address distinct practical constraints. These results establish self-optimizing multichannel optical computing as a practical platform with broad applicability beyond proof-of-concept demonstrations.

\textbf{Summary of performance across tasks.} Table~\ref{tab:performance_all} consolidates performance across all benchmarks, revealing consistent patterns that validate our architectural choices. Multichannel optical processing with random masks consistently outperforms raw pixel baselines by large margins (26--58 percentage points for classification tasks), demonstrating the fundamental value of optical feature extraction through cascaded phase transformations. RGB processing systematically outperforms grayscale processing by 5--6 percentage points when both use the same optical architecture, confirming that preserving multichannel information throughout the optical domain provides tangible benefits even without optimization. Critically, our two self-optimization strategies---Bayesian mixing (input-space) and SOC (hardware-space)---each provide additional gains of 6--7 percentage points and operate on orthogonal aspects of the system. Input mixing can be applied with fixed hardware, offering a software-level performance boost without physical reconfiguration, while SOC requires hardware-in-the-loop access but achieves comparable gains through autonomous phase mask adaptation. The fact that both low-dimensional input optimization (6 parameters) and high-dimensional hardware optimization (22,500 parameters) yield similar improvements underscores a key insight: exploiting problem structure through domain-appropriate optimization strategies can be as impactful as brute-force parameter tuning. Together, these results establish self-optimizing multichannel optical computing as a versatile platform that accommodates diverse data modalities (images and structured features) and diverse optimization strategies (gradient-free, hardware-in-the-loop).

\begin{table}[t]
\centering
\small
\caption{Performance summary across benchmarks. Classification results report accuracy; regression reports normalized RMSE (see Methods for normalization). BayesOpt: Bayesian optimization; SOC: self-organized criticality; pp: percentage points.}
\label{tab:performance_all}
\begin{tabular}{lccc}
\hline
Task / Dataset & Baseline & Optimized & Improvement \\
\hline
HAM10000 (Grayscale) & 66\% & 92\% & +26 pp \\
HAM10000 (RGB) & 67\% & 98\% & +31 pp \\
STL-10 (Grayscale) & 21\% & 77\% & +56 pp \\
STL-10 (RGB) & 25\% & 83\% & +58 pp \\
STL-10 (RGB, BayesOpt) & 83\% & 90\% & +7 pp \\
Flowers-17 (random) & 37\% & 74\% & +37 pp \\
Flowers-17 (SOC, $\sim$75 iter) & 74\% & 80\% & +6 pp \\
Abalone (nRMSE) & -- & 0.08 & -- \\
\hline
\end{tabular}
\end{table}

\textbf{Self-organized criticality as an alternative to transmission-matrix methods.} Our SOC-based approach complements recent advances in self-configuring MPLCs based on transmission matrix measurement~\cite{ref37}. While transmission-matrix methods provide powerful tools for absorbing system imperfections through interferometric measurements, they require phase-stable optical systems and bidirectional optical access for holographic field reconstruction. In contrast, our avalanche-based optimization operates with unidirectional light propagation and requires only intensity measurements at the output, trading measurement complexity for exploration of a larger configuration space through stochastic avalanche dynamics. The rapid convergence we demonstrate---typically achieving meaningful performance improvements within 75 hardware-in-the-loop iterations for our Oxford Flowers-17 experiment (Fig.~\ref{fig:soc})---suggests that SOC's multi-scale perturbations efficiently navigate the high-dimensional phase landscape despite lacking gradient information. Importantly, the two approaches address complementary deployment scenarios: transmission-matrix methods excel when interferometric stability can be maintained and fast convergence is critical, while SOC-based optimization offers advantages in environments where phase drift or vibrations make interferometric measurements challenging, or when bidirectional optical access is impractical. Future work could explore hybrid approaches combining both strategies---for instance, using transmission-matrix methods for rapid initial convergence to absorb gross misalignments, followed by SOC fine-tuning to maintain optimal performance in dynamic environments without requiring continuous phase stabilization.

\textbf{Multichannel encoding unlocks optical computing scalability.} The consistent 5--6 percentage point advantage of RGB over grayscale processing (Table~\ref{tab:performance_all}) demonstrates that preserving information dimensionality matters. This finding has direct implications for scaling optical computing to modern machine learning workloads, where high-dimensional data (hyperspectral imaging, video, multi-modal sensor fusion) is ubiquitous. Our architecture shows that optical systems need not collapse inputs to single channels, and that the interference and diffraction physics naturally exploited by cascaded phase transformations can productively mix multichannel information to generate discriminative features. Moreover, the successful application to structured numerical data (Abalone regression) indicates that multichannel encoding generalizes beyond images when appropriate spatial mappings are designed. This opens pathways to optical acceleration of tabular data processing, time-series analysis, and other non-image domains.

\textbf{Gradient-free self-optimization enables practical deployment.} A persistent challenge in optical computing is the gap between simulation-optimized designs and physical hardware. Model mismatch, calibration drift, and device nonidealities degrade the effectiveness of gradient-based training methods that assume accurate forward models. Our SOC-based optimization circumvents this issue entirely by operating directly on hardware measurements. The rapid convergence within 75 iterations (Fig.~\ref{fig:soc}) demonstrates that avalanche dynamics efficiently explore high-dimensional phase spaces without gradient information or hyperparameter tuning. This is particularly valuable for systems where differentiable wave-propagation simulators may not capture all physical effects (e.g., aberrations, dust, thermal drift). While SOC does not replace gradient-based methods---which remain powerful when accurate models are available---it provides a complementary tool for scenarios where model fidelity is uncertain or where rapid prototyping without extensive calibration is desired. The fact that Bayesian optimization of mixing coefficients (6 parameters) and SOC optimization of phase masks (22,500 parameters) achieve comparable performance gains (6--7 percentage points) highlights that exploiting problem structure, whether through low-dimensional input transformations or physics-informed heuristics, can be as impactful as brute-force parameter tuning. Together, these complementary strategies realize a truly self-optimizing optical computing platform that can autonomously adapt to task requirements at both the input and hardware levels.

\textbf{Comparison to digital and other optical baselines.} The raw pixel baselines in our experiments (Table~\ref{tab:performance_all}) use the same Ridge classifier or regressor that processes optical features, ensuring that performance gains arise from optical feature extraction rather than classifier capacity. For HAM10000 and STL-10, our results (98\% and 90\% with optimization) substantially exceed these linear baselines and approach performance levels typically achieved by convolutional neural networks on these datasets. However, direct comparison to deep learning is complicated by differences in model capacity, training data, and computational cost. The key advantage of our approach is not necessarily higher accuracy than state-of-the-art digital methods, but rather the potential for lower latency and energy consumption by performing feature extraction optically. Compared to other optical computing architectures---diffractive optical networks, reservoir computing in multimode fibers, nanophotonic circuits---our platform offers unique advantages in multichannel processing, programmability, and autonomous optimization, though systematic benchmarking across architectures remains an important direction for future work.

To contextualize our optical computing results, we compare performance to recent state-of-the-art digital methods on the same datasets (Supplementary Tables 1--2). On HAM10000, our multichannel RGB system achieves 98\% accuracy without data augmentation, matching the best recent digital methods that rely on extensive augmentation. On STL-10, our optimized system with channel mixing reaches 90\% accuracy in a fully supervised setting, placing it competitively alongside recent semi-supervised approaches that leverage unlabeled data. These comparisons demonstrate that optical feature extraction can rival sophisticated digital methods while offering potential advantages in latency and energy efficiency. The raw pixel baselines in our experiments (Table~\ref{tab:performance_all}) use the same Ridge classifier or regressor that processes optical features, ensuring that performance gains arise from optical feature extraction rather than classifier capacity. For HAM10000 and STL-10, our results substantially exceed these linear baselines and approach performance levels typically achieved by convolutional neural networks on these datasets. However, direct comparison to deep learning is complicated by differences in model capacity, training data, and computational cost. The key advantage of our approach is not necessarily higher accuracy than state-of-the-art digital methods, but rather the potential for lower latency and energy consumption by performing feature extraction optically. Compared to other optical computing architectures---diffractive optical networks, reservoir computing in multimode fibers, nanophotonic circuits---our platform offers unique advantages in multichannel processing, programmability, and autonomous optimization, though systematic benchmarking across architectures remains an important direction for future work.

\textbf{Limitations and outlook.} Our current implementation processes inputs sequentially rather than in parallel, limiting throughput. Future designs could exploit spatial or temporal multiplexing to process multiple inputs simultaneously. The digital readout layer, while simple, still requires electronic computation; fully optical classifiers using nonlinear detection or photonic circuits could eliminate this bottleneck. Our SOC implementation uses a fixed avalanche model; learned or adaptive avalanche dynamics tailored to optical propagation physics may further improve optimization efficiency. Combining input-space and hardware-space optimization simultaneously, rather than sequentially, could reveal synergistic effects not captured in our current evaluation. Finally, while we demonstrate versatility across multiple tasks, scaling to higher-resolution images, deeper cascades, or larger class counts will require addressing practical challenges such as SLM fill factor, alignment stability, and noise accumulation. Nonetheless, the core principles---multichannel encoding, phase-only cascades, and gradient-free self-optimization---provide a robust foundation for next-generation optical computing systems that can natively process the high-dimensional, multichannel data prevalent in modern machine learning applications.

\section{Methods}

\subsection*{Optical setup and system architecture}

The optical system utilizes a phase-only spatial light modulator (Holoeye PLUTO-2.1 LCOS NIR 145) illuminated by a collimated continuous-wave laser (Thorlabs, wavelength 633~nm, power 5~mW). The laser beam is expanded and collimated to uniformly illuminate a 280-pixel-wide active region on the SLM.  The beam undergoes four consecutive round-trip reflections between the SLM and a planar mirror, creating four spatially separated active regions on the SLM that function as cascaded phase masks. A slit aperture positioned before the first reflection controls beam divergence and defines the spatial extent of the propagating field. Output intensity patterns are captured by a FLIR Blackfly S camera (BFS-U3-04S2M-CS, monochrome, $720\times540$ pixels, 8-bit depth) positioned at the final detection plane.

The SLM addressability is $1920\times1080$ pixels with 8.0~$\mu$m pixel pitch. The phase masks are separated by approximately 200 pixels along the propagation axis to avoid spatial overlap. Phase values are discretized to 256 levels spanning $[0, 2\pi)$ radians, with lookup-table correction applied to linearize the phase-voltage response of the liquid-crystal modulator.

\subsection*{Input encoding and preprocessing}

\textbf{Image datasets.} STL-10 images (original resolution $96\times96$ pixels) were resized to $240\times240$ pixels and zero-padded to $280\times280$ pixels to match the SLM active area. HAM10000 images (original resolution $450\times600$ pixels) were rotated to landscape orientation ($600\times450$ pixels), resized to $240\times180$ pixels while preserving aspect ratio, and zero-padded to $280\times280$ pixels. Grayscale inputs were encoded within the central region of the SLM. For RGB inputs, the full active area was utilized, and the color channels were separated by 20 pixels in the vertical direction to reduce optical crosstalk.

\textbf{Structured numerical data.} For the Abalone regression task (4,177 samples, 8 original features), we excluded whole weight (linearly dependent on component weights) and one-hot encoded the categorical sex variable (male, female, infant), yielding 9 features. Numerical features (length, diameter, height, shucked weight, viscera weight, shell weight) were normalized to $[0, 1]$ via min-max scaling. The 9 features were spatially encoded as a $3\times3$ grid on the first SLM active region, with 30-pixel horizontal spacing and 40-pixel vertical spacing to minimize crosstalk. Each feature value was mapped to phase via linear scaling: feature value $v \in [0, 1]$ corresponds to phase $\phi = 2\pi v$.

\subsection*{Optical feature extraction and digital readout}

After propagation through the four-mask cascade, the output intensity pattern was captured by the FLIR camera. Raw camera images ($720\times540$ pixels) were cropped to the central $540\times180$ pixel region containing the output field, then downsampled by a factor of 6 via average pooling, yielding a $90\times30 = 2{,}700$-dimensional feature vector. The resulting feature vectors were normalized and employed as inputs to a Ridge classifier to perform the final classification.

For classification tasks (HAM10000, STL-10, Flowers-17), we used Ridge classification. For regression (Abalone), we used Ridge regression. All datasets were split 80\% training, 20\% test. 

\subsection*{Channel mixing optimization}

For RGB inputs, we parameterized channel mixing via six coefficients $\{c_1, c_2, c_3, c_4, c_5, c_6\}$ that define linear combinations:
\begin{itemize}
\item Mixed red: $c_1 R + (1 - c_1) G$
\item Mixed green: $c_2 G + (1 - c_2) B$
\item Mixed blue: $c_3 B + (1 - c_3) R$
\item Cross-term 1: $c_4 R + (1 - c_4) G$
\item Cross-term 2: $c_5 G + (1 - c_5) B$
\item Cross-term 3: $c_6 B + (1 - c_6) R$
\end{itemize}

Each coefficient was constrained to the interval $[0,1]$. The six-dimensional coefficient vector was optimized using Bayesian optimization, with 10\% of the data used for optimization.

\subsection*{Phase mask optimization via self-organized criticality}

We represent the nine phase masks as a single concatenated lattice of $150 \times 150$ sites, with each $50 \times 50$ mask occupying a distinct region. An Abelian sandpile grid of the same size is simulated to generate perturbation locations. At each iteration, grains are randomly added to grid sites until the local grain count exceeds the critical threshold $h_{\text{critical}} = 4$. When a site exceeds this threshold, it topples, redistributing one grain to each of its four nearest neighbors and reducing its own count by four. Cascading topplings continue until all sites stabilize below the threshold, producing an avalanche map—a binary matrix that identifies which lattice sites are selected for perturbation.

These perturbation locations are then mapped onto the corresponding pixels of the phase masks, where the actual phase updates are applied. Each selected pixel is modified by adding a random phase change drawn from a normal distribution whose scale is proportional to the perturbation magnitude. The updated phase masks are evaluated experimentally: they are uploaded to the SLM, output intensity patterns are captured for 50\% of the data, features are extracted, and test accuracy is computed. If the accuracy improves on this subset, the remaining data are also evaluated; if performance increases, the update is accepted. Otherwise, the previous mask configuration is retained. This two-stage approach prevents wasted computation on ineffective changes.  The procedure is repeated for 75 iterations on the \textit{Flowers-17} dataset, and the optimized phase masks are used for final evaluation on the complete dataset.

\subsection*{Performance metrics and normalization}
Classification accuracy is reported as the fraction of correctly predicted labels. For Abalone regression, we report normalized root mean square error (nRMSE):
\begin{equation}
\text{nRMSE} = \frac{\text{RMSE}}{y_{\text{max}} - y_{\text{min}}}
\end{equation}
where $\text{RMSE} = \sqrt{\text{mean}((y_{\text{pred}} - y_{\text{true}})^2)}$, and $y_{\text{max}}$, $y_{\text{min}}$ are the maximum and minimum target values in the training set. This normalization enables comparisons across datasets with different target ranges.

Linear discriminant analysis (LDA) projections for visualization were computed using scikit-learn's \texttt{LinearDiscriminantAnalysis} with the top 3 components retained. 

\subsubsection*{Data availability}
Any additional data generated during experiments may be obtained from the authors upon reasonable request. Datasets used in the study are publicly available through their respective access links.

\subsubsection*{Code availability}
Codes related to the results in this work publicly are available at Zenodo ~\cite{kilinc_utegin-lptselfopt_2026}.

\subsubsection*{Acknowledgments}
This work is supported by the Scientific and Technological Research Council of Turkey (T\"{U}B\.{I}TAK) under grant number 122C150.

\subsubsection*{Author contributions}
F.N.K. constructed the experimental setup and conducted numerical and experimental studies. U.T. provided advice, and was in charge of supervision. All authors participated in the analysis of results and the writing of the manuscript.

\subsubsection*{Competing interests}
All other authors declare no competing interests.

\bibliographystyle{unsrt} 
\bibliography{references}

\end{document}